\documentclass[conference]{IEEEtran}
\IEEEoverridecommandlockouts
\usepackage{cite}
\usepackage{amsmath,amssymb,amsfonts}
\usepackage{algorithmic}
\usepackage{graphicx}
\usepackage{textcomp}
\usepackage{xcolor}

\usepackage{graphicx}
\usepackage{caption}

\usepackage{pifont}  
\usepackage{booktabs} 
\usepackage{float}

\newcommand{\cmark}{\textcolor{green}{\ding{51}}}  
\newcommand{\xmark}{\textcolor{red}{\ding{55}}}    

\def\BibTeX{{\rm B\kern-.05em{\sc i\kern-.025em b}\kern-.08em
    T\kern-.1667em\lower.7ex\hbox{E}\kern-.125emX}}
    
\begin{document}

\title{Competitive Audio-Language Models with Data-Efficient Single-Stage Training on Public Data}


\author{\IEEEauthorblockN{Gokul Karthik Kumar\quad Rishabh Saraf\quad Ludovick Lepauloux\\Abdul Muneer\quad Billel Mokeddem\quad Hakim Hacid}
\IEEEauthorblockA{
\textit{Technology Innovation Institute, Abu Dhabi, UAE }\\
\texttt{\{gokul.kumar, rishabh.saraf, ludovick.lepauloux\}@tii.ae}
}}



\maketitle

\begin{abstract}
Large language models (LLMs) have transformed NLP, yet their integration with audio remains underexplored despite audio’s centrality to human communication. We introduce Falcon3-Audio, a family of Audio-Language Models (ALMs) built on instruction-tuned LLMs and Whisper encoders. Using a remarkably small amount of public audio data, less than 30K hours (5K unique), Falcon3-Audio-7B matches the best reported performance among open-weight models on the MMAU benchmark, with a score of 64.14, matching R1-AQA, while distinguishing itself through superior data and parameter efficiency, single-stage training, and transparency. Notably, our smallest 1B model remains competitive with larger open models ranging from 2B to 13B parameters. Through extensive ablations, we find that common complexities such as curriculum learning, multiple audio encoders, and intricate cross-attention connectors are not required for strong performance, even compared to models trained on over 500K hours of data.
\end{abstract}


\section{Introduction}

Large language models (LLMs) like GPT-4 \cite{achiam2023gpt}, Gemini 1.5 \cite{team2024gemini}, LLaMA 3 \cite{grattafiori2024llama3herdmodels}, Falcon 3 \cite{Falcon3}, Qwen2 \cite{yang2024qwen2}, and Phi-3 \cite{abdin2024phi} have demonstrated advanced capabilities in text understanding, reasoning, and generation. With enhancements in instruction tuning and dataset curation techniques \cite{vicuna2023, wang2023openchat, liu2024dora}, their capabilities continue to grow. However, the next frontier is multimodal intelligence, extending these powerful models to rich modalities like audio.

Audio is central to human communication, conveying not only linguistic content but also paralinguistic information like tone, emotion, and speaker intent. Enabling LLMs to understand audio could unlock powerful applications in assistive AI, conversational agents, sound event detection, and music reasoning. However, audio-language modeling remains underexplored compared to vision-language modeling, and presents unique challenges: (1) lack of established design principles for audio-language models (ALMs), (2) high diversity across audio domains (speech, music, environmental sounds), (3) high temporal resolution of audio signals, and (4) variable input lengths that complicate alignment with LLM input sequences.

\begin{figure}[t]
\centering
\includegraphics[width=0.6\columnwidth]{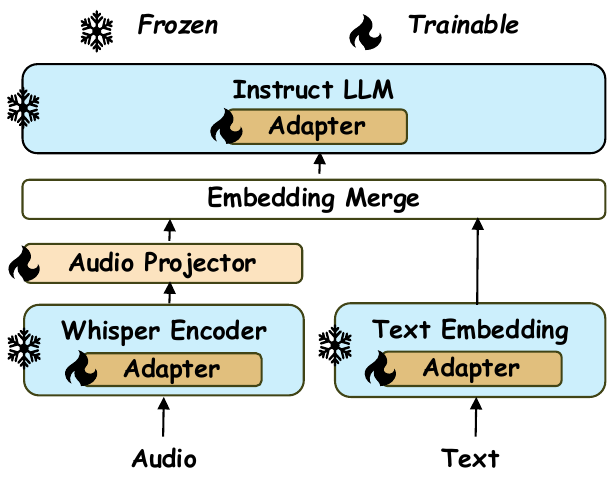}
\caption{Architecture of Falcon3-Audio which integrates Whisper audio encoder to extract features, and projects audio tokens into the Instruct LLM input space via a learnable projector.}
\label{fig:falcon_audio_architecture}
\end{figure}

Many ALMs try to solve such challenges with increasingly complex designs—multi-stage pipelines, large proprietary datasets, or intricate cross-modal attention mechanisms. This diversity reflects a lack of consensus and creates barriers to reproducibility and community-driven progress. There is a growing need for ALMs that are not only competitive but also transparent, efficient, and fully based on public resources.

\textbf{Our Contribution:} We present \textbf{Falcon3-Audio}, a family of ALMs that offer a simple, transparent, and reproducible architecture for aligning audio with instruction-tuned LLMs. Falcon3-Audio is built entirely on public data, with \textit{models of 1B, 3B, and 7B parameters and trained end-to-end via a single-stage fine-tuning strategy}. Through a systematic exploration of critical design choices, \emph{aiming for a minimalist approach}, we show that strong audio-language performance does not require complex architectures, proprietary data, or multi-stage training.

\begin{itemize}
    \item \textbf{Efficient Architecture:} Falcon3-Audio connects Whisper audio encoder to an instruction-tuned LLM through a lightweight projection module.
    
    \item \textbf{Strong Results with Public Data:} Falcon3-Audio-7B matches state-of-the-art performance among open-weight models on the MMAU benchmark, despite using less than 30K hours of audio - significantly less than similar performing models. It also achieves strong performance on AIR-Bench Foundational and Chat.
    
    \item \textbf{Transparent and Reproducible Design:} Falcon3-Audio uses only public datasets and open-source models, making it easy to reproduce, extend, and evaluate.
    
    \item \textbf{Comprehensive Ablation Studies:} We conduct detailed ablations across design factors—audio encoder, sequence length, projection design, training strategy, and data mixture—providing practical guidance for future ALMs.
\end{itemize}

\section{Related Work}

Building upon the remarkable progress in LLMs, our work on Falcon3-Audio is deeply informed by advancements in both vision-language models and the rapidly evolving audio-language models.

\subsection{Vision Language Models: Insights for Multimodal Integration}

Vision-language models (VLMs) have pioneered the integration of non-textual modalities into LLMs, achieving impressive results in tasks such as visual question answering, image captioning, and multimodal reasoning. Models like Flamingo \cite{alayrac2022flamingo}, LLaVA \cite{liu2024improvedbaselinesvisualinstruction}, ShareGPT4V \cite{chen2025sharegpt4v}, Qwen-Vision \cite{bai2023qwen}, and IDEFICS \cite{laurençon2023obelicsopenwebscalefiltered} demonstrate effective strategies for combining image and text inputs. These models commonly leverage powerful pre-trained unimodal encoders—such as CLIP \cite{radford2021learningtransferablevisualmodels} or SigLIP \cite{zhai2023sigmoidlosslanguageimage} for vision, and LLaMA or Mistral for language—integrating them through cross-attention or projection-based mechanisms. VLMs often interleave image features within the LLM architecture using cross-attention, or directly inject visual embeddings by concatenating patch-wise image representations with text tokens.  

Crucially, the VLM field offers valuable architectural patterns and integration techniques applicable to other modalities. For instance, the concept of treating frame-wise spectrogram features from audio encoders (like Whisper) as analogous to patch-wise image embeddings in VLMs provides a foundational parallel for Audio-Language Model design. While both VLMs and ALMs face challenges in feature alignment and scaling, audio introduces unique complexities such as variable input lengths and the diversity of audio domains (speech, music, environmental sounds). The progress in VLMs, particularly in efficient feature integration and projection, serves as a crucial stepping stone and inspiration for extending LLMs to the audio modality, informing our approach in Falcon3-Audio.

\subsection{Audio Language Models: Towards General Audio Understanding}

The field of Audio-Language Models (ALMs) is rapidly emerging as a critical area of research, aiming to equip LLMs with the ability to understand and reason with audio. Early ALM research focused on cross-modal encoders like AudioCLIP \cite{guzhov2022audioclip} and CLAP \cite{elizalde2023clap}, primarily for tasks like audio retrieval and classification by aligning audio and text representations in a shared embedding space. The advent of powerful LLMs has shifted the paradigm towards directly integrating audio processing within LLMs, unlocking more sophisticated audio-grounded reasoning and generation capabilities. This has led to specialized ALMs like MuLLama \cite{liu2024music} and MusiLingo \cite{deng2023musilingo} for music understanding, and LTU \cite{gong2024listenthinkunderstand} for non-speech audio tasks. SLAM-ASR \cite{ma2024embarrassingly} is a speech recognition-focused effort, complementary to our work. Recent advances in acoustic tokenization, such as WavTokenizer \cite{ji2024wavtokenizer}, have enabled efficient compression of audio signals into discrete tokens, facilitating scalable speech generation with large language model backbones. These early LLM-based ALMs demonstrated the potential of audio-text alignment but often remained limited in scope and generalization.

More recent advancements have focused on developing general-purpose ALMs capable of handling a wider range of audio tasks. Models like Pengi \cite{deshmukh2023pengi}, LTU-AS \cite{10389742}, SALMONN \cite{tang2024salmonngenerichearingabilities}, GAMA \cite{ghosh2024gama}, Qwen-Audio \cite{chu2023qwenaudioadvancinguniversalaudio}, and Qwen2-Audio \cite{chu2024qwen2audiotechnicalreport} represent this trend, integrating diverse audio types (speech, environmental sounds, music) into LLM frameworks. These models use various techniques: LTU-AS replaced AST encoders with Whisper and incorporated time/layer transformers; SALMONN uses cross-attention to fuse features from multiple encoders like Whisper and BEATs \cite{chen2022beatsaudiopretrainingacoustic}; GAMA utilizes multi-layer transformer aggregators; and the Qwen-Audio series has progressively refined training frameworks and data scale. Qwen2-Audio, in particular, streamlined training with natural language prompts and expanded to multi-turn audio interactions.

\textbf{Concurrent Works:} Several concurrent works—such as R1-AQA \cite{li2025reinforcement}, Audio Flamingo 2 \cite{ghosh2025audio}, Phi-4 Mini \cite{abouelenin2025phi}, and Qwen2.5 Omni \cite{xu2025qwen2}—were released after the completion of our experiments. While relevant, they are excluded from some of our evaluation tables due to incomplete reporting: none provide results on AIR-Bench Foundational \cite{yang2024airbenchbenchmarkinglargeaudiolanguage}, and only Phi-4 Mini reports on AIR-Bench Chat \cite{yang2024airbenchbenchmarkinglargeaudiolanguage}. For MMAU \cite{sakshi2024mmau}, Qwen2.5 Omni reports only on the test-mini split.
\textbf{R1-AQA} fine-tunes Qwen2-Audio Instruct using Group Relative Policy Optimization (GRPO) on 38K task-aligned samples. It inherits Qwen2-Audio Instruct’s dependency on $>$500K hours of non-public data, a large 8.4B model, and a multi-stage training pipeline—with GRPO introducing an additional fine-tuning stage.
\textbf{Audio Flamingo 2}, while transparent, uses a more complex setup: a separately pretrained CLAP \cite{elizalde2023clap} encoder is fused into a 3B LLM via cross-attention layers within intermediate transformer blocks. This requires careful finetuning to preserve the LLM’s language capabilities. Falcon3-Audio, in contrast, employs a simple projection into the LLM’s input embeddings and trains end-to-end in a single stage.
\textbf{Phi-4 Mini} is a 5.6B model trained on over 2 million hours of private speech data, in addition to non-speech audio, images, videos, and text. It uses a multi-stage curriculum with vision and audio adapters, but provides limited transparency in terms of training data sources and fine-tuning details.
\textbf{Qwen2.5 Omni} also incorporates a vision encoder, like Phi-4 Mini, but uses a larger 10.7B model. It is trained on considerably more data than Falcon3-Audio—sourced from undisclosed datasets—and follows a multi-stage curriculum. The lack of transparency in both data and training pipeline limits reproducibility.

\textbf{Despite these recent advancements}, many of these models rely on complex architectures and large-scale proprietary data. Falcon3-Audio, by contrast, achieves strong performance using a simpler, transparent setup based solely on public resources. Full performance comparisons are presented in Section~\ref{sec:performance}.

\section{Design Choices: Guiding Principles for Falcon3-Audio}

To realize Falcon3-Audio as an efficient, performant, and accessible audio-language model, we conducted a systematic exploration of key design elements. This section details the critical considerations that shaped our final model.  
Our goal was to identify a configuration that optimally balances simplicity, computational efficiency, and state-of-the-art performance. 
Detailed results of our ablation studies are presented in the supplementary material (Section I). 

\subsection{Language Model Foundation: Base or Instruct-Tuned?}

For Falcon3-Audio's language backbone, we chose a Falcon3 \cite{Falcon3} family of instruction-tuned LLMs, available in 1B, 3B, and 7B variants, which are established strong performers within their respective size categories. For the 7B variant, we used an internal version trained with less code and math data compared to the released model, as we observed improved performance with this configuration. Our decision to utilize instruction-tuned models, rather than base model counterparts, was driven by the hypothesis that their inherent instruction-following capabilities would be advantageous for audio-language tasks, often involving complex instructions and queries. Empirical validation confirmed this hypothesis.

\subsection{Audio Encoder Selection: Capturing the Spectrum of Audio Information}

Selecting an appropriate audio encoder is crucial for extracting audio features that effectively generalize across the diverse landscape of audio tasks.   While options like AST \cite{gong2021ast}, CLAP \cite{elizalde2023clap}, ImageBind Audio Encoder \cite{Girdhar_2023_CVPR}, and BEATs \cite{chen2022beatsaudiopretrainingacoustic} each offer unique strengths, we prioritized Whisper \cite{radford2023robust} for Falcon3-Audio.  Whisper has emerged as a leading encoder, particularly for speech-related tasks, and has demonstrated strong performance even in non-speech audio domains \cite{Gong_2023, chu2024qwen2audiotechnicalreport}.  Given the importance of speech understanding and Whisper's overall robustness, we used the Whisper Medium English encoder for our 7B/3B models, and the Whisper Small English encoder for 1B, balancing performance and parameter efficiency. Ablation studies confirmed that Whisper Medium (English) provided the best performance for our 7B model.

\subsection{Audio Sequence Handling: Balancing Efficiency and Temporal Resolution}

Audio signals, characterized by high temporal density, pose a challenge for efficient processing. Even Whisper's downsampling to 50 tokens per second results in a much higher token rate compared to text. 
Approaches to audio sequence reduction vary. LTU-AS \cite{10389742} uses time and layer transformers for aggressive reduction. Shuka-v1 \cite{sarvam2024shuka} utilizes stacking from SLAM-ASR \cite{ma2024embarrassingly} to reduce sequence length.
Based on ablation studies, we found that aggressive sequence length reduction was not necessary for our architecture and tasks. 
Therefore, we adopted a straightforward pooling layer, similar to Qwen2-Audio \cite{chu2024qwen2audiotechnicalreport}, which reduces the token rate to 25 tokens per second. This provides a computationally efficient solution. 

\subsection{Connector Architecture: Bridging Audio and Text Embeddings}

The architecture connecting the audio encoder and the language model is a critical design choice. We followed a projection-based approach, proven successful in VLMs like LLaVA-v1.5 \cite{liu2024improved} and ALMs like Qwen2-Audio \cite{chu2024qwen2audiotechnicalreport}. Falcon3-Audio uses a learnable projector to map the audio encoder's output dimensionality to the LLM's input dimension. This projector is augmented with an activation function, an additional linear layer, and layer normalization, inspired by Shuka-v1 \cite{sarvam2024shuka}. Ablation studies showed that incorporating intermediate layer features from the Whisper encoder did not improve performance, so we opted for the simpler final-layer projection. 

\subsection{Training Data Composition:  Ensuring Generalization Across Audio Domains}
\label{sec:data-mixture}

The selection and composition of training data are paramount for developing ALMs capable of generalizing to a wide range of audio-related tasks.
For Falcon3-Audio, we adopted the Open-ASQA dataset from LTU-AS \cite{10389742} as our primary training corpus. Open-ASQA, the largest publicly available audio instruction dataset at the time of our development, comprises approximately 10 million samples (26K hours) drawn from diverse sources, including AudioSet \cite{gemmeke2017audio}, AudioCaps \cite{kim2019audiocaps}, FSD50K \cite{fonseca2021fsd50k}, FreeSound \cite{fonseca2017freesound}, LibriTTS \cite{zen2019libritts}, VGGSound \cite{chen2020vggsound}, and VoxCeleb \cite{nagrani2017voxceleb}. To further enhance voice interaction capabilities, we supplemented this with 440K samples (approximately 1K hours) from a synthetic voice-instruction dataset \cite{InstructionSpeech2024}. Ablation studies indicated that adding further datasets did not improve performance on our target benchmark. 


\subsection{Training Strategy: Simplicity vs. Curriculum Learning}

In contrast to complex, multi-staged curriculum learning approaches, we investigated a simpler \emph{single-stage end-to-end fine-tuning} strategy.  This strategy trains the entire model on the full task spectrum from the outset. 
Ablation studies showed that this single-stage approach was superior to a two-stage curriculum for our model. Furthermore, we found that training all modules jointly (audio encoder, projector, and LLM) yielded the best results. 



\begin{table*}[t]
\centering
\caption{MMAU Benchmark (Test) results. TDTS: Transparent Data and Training Strategy; Train Data: audio hours. \textbf{Top scores} are bolded; \underline{second-best} are underlined; values are left empty where unavailable; \textsuperscript{+}indicates concurrent work. Sources (except Falcon3-Audio): \cite{sakshi2024mmau, abouelenin2025phi}.}
\begin{tabular}{p{0.5cm} p{3.5cm} p{1cm} p{1.4cm} p{1.4cm} p{1cm} p{1cm} p{1cm} p{1cm}}
\toprule
\textbf{Rank} & \textbf{Method} & \textbf{TDTS} & \textbf{Model Size} & \textbf{Train Data} & \textbf{Average} & \textbf{Sound} & \textbf{Music} & \textbf{Speech} \\
\midrule

\multicolumn{9}{c}{\emph{End-To-End ALMs}} \\
\midrule
1  & R1-AQA\textsuperscript{+} & \xmark & 8.4B & 500K-2M & \textbf{64.36} & \underline{69.76} & \underline{61.40} & \textbf{62.70} \\
2  & \emph{Falcon3-Audio 7B} & \textbf{\cmark} & 7.8B & 0-30K & \underline{64.14} & \textbf{71.27} & 58.53 & \underline{62.63} \\
3  & Audio Flamingo 2\textsuperscript{+} & \cmark & 4.7B & 30K-100K & 59.42 & 65.10 & \textbf{72.90 }& 40.26 \\
4  & \emph{Falcon3-Audio 3B} & \textbf{\cmark} & 3.6B & 0-30K & 57.96 & 67.33 & 52.07 & 54.47 \\
5  & Phi-4-Mini\textsuperscript{+} & \xmark & 5.6B & 2M+ & 55.56 & & &  \\
6  & Qwen2-Audio Instruct & \xmark & 8.4B & 500K-2M & 52.50 & 45.90 & 53.26 & 45.90 \\
7  & \emph{Falcon3-Audio 1B} & \textbf{\cmark} & 1.8B & 0-30K & 50.80 & 59.90 & 46.30 & 46.20 \\
8  & Qwen-Audio Chat & \xmark & 8.4B & 100K-500K & 41.86 & 56.73 & 40.90 & 27.95 \\
9  & SALMONN & \cmark & 13B & 0-30K & 32.77 & 40.30 & 33.76 & 24.24 \\
10  & GAMA & \cmark & 7B &   & 31.81 & 45.40 & 30.83 & 19.21 \\
11  & MuLLaMA & \cmark & 7B & & 30.66 & 44.80 & 30.63 & 16.56 \\
12  & GAMA-IT & \cmark & 7B &  & 29.02 & 43.23 & 28.00 & 15.84 \\
13  & LTU-AS & \cmark & 7B & 0-30K & 18.90 & 24.96 & 10.46 & 21.30 \\
14 & Audio Flamingo Chat & \cmark & 2.2B & 0-30K  & 18.87 & 28.26 & 18.20 & 10.16 \\
15 & LTU & \cmark & 7B & 0-30K & 18.51 & 25.86 & 12.83 & 16.37 \\
16 & MusiLingo & \cmark & 7B & & 13.39 & 27.76 & 6.00 & 6.42 \\
17 & M2UGen & \cmark & 7B &  & 12.87 & 3.69 & 30.40 & 4.53 \\
\midrule
\multicolumn{9}{c}{\emph{Baselines}} \\
\midrule
 & \multicolumn{4}{l}{Human (Test-Mini)}& 82.23 & 86.31 & 78.22 & 82.17 \\
 & \multicolumn{4}{l}{Random Guess} & 25.92 & 25.73 & 26.53 & 25.5 \\
 & \multicolumn{4}{l}{Most Frequent Choice} & 26.5 & 25.73 & 23.73 & 30.33 \\
\midrule
\multicolumn{9}{c}{\emph{APIs And Cascaded Approaches}} \\
\midrule
 & \multicolumn{4}{l}{Qwen2-Audio Instruct + GPT4o} & 58.74 & 55.83 & 51.73 & 68.66 \\
 & \multicolumn{4}{l}{Qwen2-Audio Instruct + Llama-3-Instruct} & 53.57 & 49.10 & 62.70 & 55.25 \\
 & \multicolumn{4}{l}{Gemini Pro v1.5} & 52.97 & 54.46 & 48.56 & 55.90 \\
 & \multicolumn{4}{l}{EnCLAP/MuLLaMA/Whisper + GPT4o}& 48.65 & 35.80 & 39.52 & 68.27 \\
 & \multicolumn{4}{l}{EnCLAP/MuLLaMA/Whisper + Llama-3-Instruct} & 45.87 & 33.73 & 42.36 & 61.54 \\
\bottomrule
\end{tabular}
\label{tab:mmau_results_test}
\end{table*}

\begin{table*}[t]
\centering
\caption{AIR-Bench Foundational Benchmark results. \textbf{Top scores} are bolded; \underline{second-best} are underlined.}
\begin{tabular}{lcccc}
\toprule
\textbf{Task} & \textbf{\emph{Falcon3-Audio 7B}} & \textbf{Qwen2-Audio Instruct} & \textbf{\emph{Falcon3-Audio 3B}} & \textbf{\emph{Falcon3-Audio 1B}} \\
\midrule
\emph{Rank} & 1 & 2 & 3 & 4 \\
\midrule
\emph{Total Average} & \textbf{54.00} & \underline{44.00} & 42.00 & 38.00 \\
\midrule
\emph{Sound Average} &  \textbf{65.2}  &  49.8  &  \underline{53.4}  &  47.7  \\
\midrule
Audio grounding & \textbf{70.70} & 17.80 & \underline{60.00} & 42.10 \\
Vocal sound classification & \textbf{90.70} & 71.10 & \underline{74.90} & 72.10 \\
Acoustic scene classification & \textbf{43.60} & 40.50 & \underline{40.80} & 35.20 \\
Sound question answering & \textbf{71.60} & \underline{62.80} & 52.40 & 50.40 \\
\midrule
\emph{Music Average} &  \textbf{50.5}  &  \underline{46.1}  &  42.2  &  36.4  \\
\midrule
Music instruments classification & \textbf{58.50} & \underline{49.60} & 46.90 & 39.50 \\
Music genre classification & \textbf{65.10} & \underline{63.90} & 59.90 & 50.60 \\
Music note analysis-pitch & \textbf{28.80} & \underline{24.30} & 19.60 & 23.90 \\
Music note analysis-velocity & \textbf{25.00} & \underline{24.70} & 22.80 & 19.10 \\
Music question answering & \textbf{58.70} & \underline{56.00} & 41.80 & 38.80 \\
Music emotion detection & \textbf{45.70} & 38.70 & \underline{39.90} & 29.50 \\
\midrule
\emph{Speech Average} &  \textbf{50.4}  &  \underline{43.5}  &  35.1  &  34.2  \\
\midrule
Speech grounding & 22.20 & \underline{26.30} & 20.30 & \textbf{26.90} \\
Spoken language identification & \textbf{66.00} & \underline{38.10} & 4.90 & 13.70 \\
Speaker gender recognition & \textbf{58.50} & \underline{52.50} & 23.90 & 29.60 \\
Emotion recognition & \textbf{64.10} & 35.40 & \underline{60.30} & 52.60 \\
Speaker age prediction & \underline{24.20} & 22.30 & \textbf{26.40} & 8.30 \\
Speech entity recognition & 36.90 & \textbf{48.30} & 26.20 & \underline{46.90} \\
Intent classification & \underline{71.70} & \textbf{78.00} & 51.40 & 44.70 \\
Speaker number verification & \underline{38.30} & 38.00 & \textbf{39.20} & 21.10 \\
Synthesized voice detection & \underline{49.80} & \textbf{51.90} & 49.30 & 49.70 \\
\bottomrule
\end{tabular}
\label{tab:air_bench_foundational_results}
\end{table*}

\begin{table*}[t]
\centering
\caption{AIR-Bench Chat Benchmark results. Train Data: audio hours . \textbf{Top scores} are bolded; \underline{second-best} are underlined; values are left empty where unavailable; \textsuperscript{+}indicates concurrent work. Sources (except Falcon3-Audio): \cite{chen2022beatsaudiopretrainingacoustic, abouelenin2025phi}.}
\begin{tabular}{p{0.5cm} p{3.5cm} p{1.4cm} p{1.4cm} p{1cm} p{1cm} p{1cm} p{1cm} p{1.2cm}}
\toprule
\textbf{Rank} & \textbf{Model} & \textbf{Model Size} & \textbf{Train Data} & \textbf{Average} & \textbf{Speech} & \textbf{Sound} & \textbf{Music} & \textbf{Mixed}\\ 
\midrule
\multicolumn{9}{c}{\emph{End-To-End ALMs}} \\
\midrule
1 & Phi-4-Mini\textsuperscript{+} & 5.6B & 2M+ & \textbf{6.98} &  & & & \\
2 & Qwen2-Audio Instruct & 8.4B & 500K-2M & \underline{6.93} & \textbf{7.18} & \textbf{6.99 }& \textbf{6.79} & \textbf{6.77} \\
3  & \emph{Falcon3-Audio 7B}  & 7.8B & 0-30K & 6.20 & \underline{6.93} & 6.74 & 5.64 & 5.48 \\
4 & SALMONN  & 13B & 0-30K & 6.11 & 6.16 & 6.28 & \underline{5.95} & \underline{6.08} \\
5 & Qwen-Audio Chat & 8.4B & 100K-500K & 6.08 & 6.47 & \underline{6.95} & 5.52 & 5.38 \\
6  & \emph{Falcon3-Audio 3B} & 3.6B & 0-30K &  
5.88 & 6.65 & 6.34 & 5.37 & 5.17 \\
7  & \emph{Falcon3-Audio 1B} & 1.8B & 0-30K &
5.67 & 6.33 & 6.14 & 5.48 & 4.74 \\
8 & BLSP & 7B &  & 5.33 & 6.17 & 5.55 & 5.08 & 4.52 \\
9 & Pandagpt & 7B &  & 4.25 & 3.58 & 5.46 & 5.06 & 2.93 \\
10 & Next-gpt & 8.2B &  & 4.13 & 3.86 & 4.76 & 4.18 & 2.92 \\
11 & SpeechGPT & 13B &  & 1.15 & 1.57 & 0.95 & 0.95 & 1.14 \\
12 & Macaw-LLM & 7B &  & 1.01 & 0.97 & 1.01 & 0.91 & 1.00 \\
\midrule
\multicolumn{9}{c}{\emph{APIs}} \\
\midrule
  & \multicolumn{3}{l}{Qwen-Audio-Turbo} & 6.34 & 7.04 & 6.59 & 5.98 & 5.77 \\
  & \multicolumn{3}{l}{Gemini-1.5-pro} & 5.70 & 6.97 & 5.49 & 5.06 & 5.27 \\
\bottomrule
\end{tabular}
\label{tab:air_bench_results}
\end{table*}

\section{Methodology}

Falcon3-Audio integrates a pretrained Whisper audio encoder, a projection module, and an Instruct-tuned LLM for efficient audio-language understanding (Figure~\ref{fig:falcon_audio_architecture}). We describe each component and our single-stage end-to-end training strategy below.  

\subsection{Falcon3-Audio Architecture}

\subsubsection{Audio Encoder}
Falcon3-Audio uses the Whisper model \cite{radford2023robust} to extract audio features. We use Whisper Medium (English) for the 7B and 3B variants and Whisper Small (English) for the 1B variant.  

Given an audio sequence $x$ of length $T$, the Whisper encoder produces a feature representation $h_{\text{audio}}$ through a series of $L$ layers:
\begin{equation}
h^{(l)} = f_{\text{AE}}(h^{(l-1)}; \theta_{\text{AE}}^{(l)}), \quad h^{(0)} = x
\end{equation}
\begin{equation}
h_{\text{audio}} = h^{(L)}
\end{equation}
where $h^{(l)}$ is the feature at layer $l$, $f_{\text{AE}}$ is the transformation, and $\theta_{\text{AE}}^{(l)}$ are the parameters. 

\subsubsection{Projection Module}
A lightweight projection module aligns $h_{\text{audio}}$ with the LLM's embedding space. It consists of: an initial layer normalization, a linear layer ($W_1, b_1$), GELU activation ($\sigma$), a second linear layer ($W_2, b_2$), and a final layer normalization: 

\begin{align}
    h_{\text{norm1}} &= \text{LayerNorm}(h_{\text{audio}}) \\
    h_{\text{proj1}} &= W_1 h_{\text{norm1}} + b_1 \\
    h_{\text{act}} &= \sigma(h_{\text{proj1}}) \\
    h_{\text{proj2}} &= W_2 h_{\text{act}} + b_2 \\
    h_{\text{final}} &= \text{LayerNorm}(h_{\text{proj2}})
\end{align}

\subsubsection{Large Language Model}
Falcon3-Audio uses Falcon3 Instruct LLMs as its core. Given a tokenized text input $z$ with embedding $\text{EMB}(z)$, Falcon3-Audio processes the merged sequence:
\begin{equation}
h_{\text{input}} = [h_{\text{final}}, \text{EMB}(z)]
\end{equation}
The LLM then generates a probability distribution over text tokens:
\begin{equation}
p(y | x, z) = f_{\text{LLM}}(h_{\text{input}}; \theta_{\text{LLM}})
\end{equation}
where $f_{\text{LLM}}$ is the LLM and $\theta_{\text{LLM}}$ are its parameters. 

\subsection{Single-Stage End-to-End Training}

Falcon3-Audio is trained end-to-end, maximizing the likelihood of generating correct completion tokens. The training dataset, $\mathcal{D}_{\text{Train}}$, combines Open-ASQA \cite{10389742} (approximately 10 million samples, 26K hours) and a synthetic voice instruction dataset (440K samples, 1K hours):
\[
\mathcal{D}_{\text{Train}} =  \mathcal{D}_{\text{Open-ASQA}} \cup \mathcal{D}_{\text{Voice-Instruction}}
\]

We fine-tune the projection module, audio encoder, and LLM using Low-Rank Adaptation (LoRA). The training objective is the negative log-likelihood:
\[
 -\sum_{(x, z, y) \in \mathcal{D}_{\text{Train}}} \log p(y | x, z; \theta_{\text{Projector}}, \theta_{\text{AE}}^{LoRA}, \theta_{\text{LLM}}^{LoRA})
\]
where $\theta_{\text{Projector}}$, $\theta_{\text{AE}}^{LoRA}$, and $\theta_{\text{LLM}}^{LoRA}$ are the parameters of the projection module and the LoRA-adapted parameters of the audio encoder and LLM, respectively. This approach ensures robust audio-text alignment and efficient use of public data.

\section{Experimental Evaluation}


\subsection{Experimental Setup}


\textbf{Model Variants:} To evaluate scalability and deployment flexibility, we trained Falcon3-Audio in three model sizes: 7B, 3B, and 1B parameters. Each variant shares a consistent architecture. Table \ref{tab:falcon3_audio_params} provides a detailed parameter breakdown for each Falcon3-Audio model size.

\textbf{Training Infrastructure:}  Our training pipeline was built using the Qwen2AudioForConditionalGeneration class from \emph{HuggingFace Transformers}. We executed distributed training via \emph{HuggingFace Accelerate} with DeepSpeed ZeRO-2 optimization and the \emph{HuggingFace TRL}'s SFT Trainer.  Total training time was approximately 20 hours.

\textbf{Compute Resources:}  Falcon3-Audio models were trained on AWS SageMaker, utilizing 8 nodes of P5 instance types, each equipped with 8$\times$80GB H100 GPUs.



\textbf{Hyperparameter Settings} and the \textbf{Prompt Format} are shown in the supplementary material (Section II). 

\begin{table}[t]
\centering
\caption{Parameter breakdown of Falcon3-Audio models.}
\begin{tabular}{c c c c c}
\toprule
\textbf{Variant} & \textbf{AE} & \textbf{Proj.} & \textbf{LLM} & \textbf{Total} \\ \midrule
\textbf{7B}& 307M& 12M& 7.4B& 7.8B\\
\textbf{3B}& 307M& 12M& 3.2B& 3.6B\\
\textbf{1B}& 88M& 5M& 1.7B& 1.8B\\ \bottomrule
\end{tabular}
\label{tab:falcon3_audio_params}
\end{table}

\subsection{Performance Evaluation} 
\label{sec:performance}

We evaluated Falcon3-Audio on the MMAU benchmark (10K samples) \cite{sakshi2024mmau} for multi-domain audio understanding, and the AIR-Bench Foundational (22K samples) and AIR-Bench Chat (2.2K samples) benchmarks \cite{yang2024airbenchbenchmarkinglargeaudiolanguage} for task-specific and open-ended audio understanding, respectively. See the supplementary material (Section III) for more details on these benchmarks. We fixed all random seeds (Python, NumPy, PyTorch, CUDA) to 0. On both MMAU and AIR-Bench Foundational, with a non-zero temperature, this yielded identical discrete answer strings, out of four choices per sample, across three runs.

\subsubsection{MMAU - Matching State-of-the-Art with Data Efficiency}
\label{sec:mmau}

For MMAU evaluation, we configured Falcon3-Audio by setting temperature = 0.1, top-p = 0.8, top-k = 500, and a maximum of 32 new tokens. Table~\ref{tab:mmau_results_test} summarizes results across related models.

Falcon3-Audio-7B achieves an average accuracy of 64.14\% on the MMAU test set — matching the top-performing R1-AQA, which fine-tunes Qwen2-Audio Instruct using GRPO on 38K samples. In contrast, Falcon3-Audio is trained via a significantly simpler single-stage strategy, using less than 30K hours of public audio, without any curriculum learning or GRPO-style policy optimization. Qwen2-Audio Instruct, by comparison, uses over 500K hours of undisclosed data, and R1-AQA adds an additional GRPO stage on top. Falcon3-Audio also has a smaller parameter footprint than Qwen2-Audio Instruct (8.4B), R1-AQA (8.4B), and SALMONN (13B).

Despite over 90\% overlap in training data with LTU-AS, Falcon3-Audio models demonstrate a substantial performance gain over LTU-AS’s 18.90\% average accuracy. Notably, even our smallest variant, Falcon3-Audio-1B, ranks just behind Qwen2-Audio Instruct and outperforms many larger open-weight models in the 2B–13B range. These results show that strong generalization across diverse audio domains is achievable without massive data or complex multi-stage pipelines.

\subsubsection{AIR-Bench Foundational - Clear Gains in Instructional Audio Tasks}
\label{sec:airbench_foundation}

We evaluated Falcon3-Audio on the AIR-Bench Foundational benchmark, which covers a broad set of instruction-following tasks across speech, sound, and music. To ensure fair comparison, we corrected limitations in the original scoring script, which (1) ignored invalid outputs (e.g., empty or malformed generations), and (2) checked only for answer option identifiers (e.g., ``A", ``B") without verifying the actual answer content. Our revised scorer treats invalid generations as incorrect and matches the answer content textually (see supplementary material, Section IV), similar to the MMAU evaluation.

Due to the compute budget, we compare Falcon3-Audio primarily with Qwen2-Audio Instruct, a strong baseline on both MMAU and AIR-Bench Chat. As shown in Table~\ref{tab:air_bench_foundational_results}, Falcon3-Audio-7B outperforms Qwen2-Audio Instruct by a wide margin of over 10 percentage points. Falcon3-Audio-3B also performs competitively, narrowing the gap significantly. Importantly, Falcon3-Audio was trained on a relatively modest ($<$30K hours), general-purpose dataset without tailoring for specific AIR-Bench tasks. In contrast, Qwen2-Audio Instruct benefits from multi-stage training and over 15× more data from undisclosed sources. These results highlight Falcon3-Audio’s strong instruction-following capabilities and broad generalization.

\subsubsection{AIR-Bench Chat - Competitive Results}
\label{sec:airbench_chat}

We evaluated Falcon3-Audio on the AIR-Bench Chat using GPT-4 Turbo, following the official setup (temperature 1.0 with a 50\% chance of switching to 2.0). We ran three evaluation trials and report the median score. Falcon3-Audio used the same generation settings as in Section~\ref{sec:mmau}, except with a repetition penalty of 1.1 and a maximum of 512 new tokens.

As shown in Table~\ref{tab:air_bench_results}, Falcon3-Audio-7B ranks third among end-to-end open-weight ALMs, behind Phi-4 Mini and Qwen2-Audio Instruct. Both rely on significantly larger and less transparent pipelines: Phi-4 Mini is trained on over 2M hours of multimodal data with a multi-stage curriculum, while Qwen2-Audio Instruct uses over 500K hours of data from undisclosed sources. In contrast, Falcon3-Audio is trained on $<$30K hours of public audio in a single-stage setup, yet remains competitive.

It is worth noting that Falcon3-Audio-7B outperforms both Phi-4 Mini and Qwen2-Audio Instruct by around 10 points on MMAU (Section~\ref{sec:mmau}). 
Falcon3-Audio-7B also surpasses Qwen2-Audio Instruct on AIR-Bench Foundational (Section~\ref{sec:airbench_foundation}), but ranks lower on Chat—suggesting potential bias in GPT-4-based evaluation, which may favor certain stylistic responses. See the supplementary material (Section V) for illustrative examples supporting this observation.
Nevertheless, Falcon3-Audio-1B performs strongly, surpassing several larger models (7B–13B), highlighting the efficiency of our architecture and training strategy in open-ended audio understanding

\section{Conclusion and Future Directions}

We presented Falcon3-Audio, a family of ALMs that demonstrate strong performance across diverse benchmarks through a simple, transparent, and data-efficient design built entirely on public resources. While several concurrent efforts have recently emerged, Falcon3-Audio remains among the most reproducible and data-efficient models to date. In future work, we aim to:
\begin{itemize}
    \item Expand the training corpus to include more diverse and multilingual audio data.
    \item Explore larger models and reinforcement learning-based alignment techniques.
    \item Extend the framework to multimodal settings including audio, vision, and text.
    \item Investigate instruction-driven audio synthesis.
\end{itemize}

\clearpage
\bibliographystyle{IEEEbib}
\bibliography{references}

\clearpage
\setcounter{section}{0}
\section*{Supplementary Material}

\begin{table*}[t]
\centering
\caption{Ablation studies on Falcon3-Audio 7B in the MMAU Benchmark Test-Mini Split. The base model achieves a total accuracy of 65.0, with task-wise accuracies of 69.4 (sound), 62.3 (music), and 63.0 (speech). The table reports relative accuracy changes when modifying different components.}
\begin{tabular}{l c c c c}
\toprule
\textbf{Ablation Category} & \textbf{Total Accuracy} & \textbf{Sound} & \textbf{Music} & \textbf{Speech} \\ \midrule

\multicolumn{5}{c}{\emph{Audio Encoder (Size Variants)}} \\ \midrule
Whisper Small   & -1.4 & -2.7 & -0.9 & -0.6 \\
Whisper Large  & -2.0 & -1.5 & -3.6 & -0.9 \\
Whisper Turbo  & -2.2 & -1.8 & -2.1 & -2.7 \\ \midrule

\multicolumn{5}{c}{\emph{Audio Encoder (Pretraining Data Variant)}} \\ \midrule
Whisper Medium - Multilingual& -4.5 & -3.3 & -1.8 & -8.4 \\ \midrule

\multicolumn{5}{c}{\emph{Audio Sequence Length Reduction}} \\ \midrule
Stack Factor 2  & +0.4  & -1.8  & -1.2  & +4.2 \\
Stack Factor 4  & -1.6 & -3.0  & -2.1  & +0.3 \\
Stack Factor 8  & -3.1 & -4.2  & +1.8   & -6.9 \\ \midrule

\multicolumn{5}{c}{\emph{Connector Architecture (Feature Aggregation Before Projector)}} \\ \midrule
Feature Extraction Every 12 Layers & -1.3 & -2.7 & -1.2 & 0.0 \\
Feature Extraction Every 6 Layers  & -2.9 & -2.4 & -4.2 & -2.1 \\
Feature Extraction Every 3 Layers  & -2.7 & -3.3 & -0.9 & -3.9 \\ \midrule

\multicolumn{5}{c}{\emph{Connector Architecture (Feature Aggregation After Projector)}} \\ \midrule
Feature Extraction Every 12 Layers & -2.2 & -1.2 & -1.8 & -3.6 \\
Feature Extraction Every 6 Layers   & -4.6 & -4.8 & -3.9 & -5.1 \\
Feature Extraction Every 3 Layers   & -2.4 & -2.4 & -2.7 & -2.1 \\ \midrule

\multicolumn{5}{c}{\emph{Data Mixture (Beyond Base Training Set)}} \\ \midrule
+ LibriSpeech 10K-hour Subset  & -1.6 & -4.2 & +1.5  & -2.1 \\
+ Auto-ACD   & -1.2 & -2.4 & -0.3  & -0.9 \\
+ LibriSpeech 10K-hour Subset + Auto-ACD  & -0.1 & -2.4 & +1.2  & +0.9 \\ \midrule

\multicolumn{5}{c}{\emph{Training Strategy (Curriculum Learning)}} \\ \midrule
Two-Stage Training  & -2.1 & -1.8  & -1.8  & -2.7  \\ \midrule

\multicolumn{5}{c}{\emph{Training Strategy (Trainable Modules)}} \\ \midrule
Frozen Only Audio Encoder  & -3.0  & -3.9  & -3.3  & -1.8  \\
Frozen Both Audio Encoder \& LLM  & -26.3 & -32.5 & -24.3 & -22.3 \\
Frozen Only LLM  & -27.0 & -31.9 & -28.8 & -20.5 \\
\bottomrule
\end{tabular}
\label{tab:falcon3_audio_ablation}
\end{table*}

\section{Ablation Studies} 
\label{sec:ablation_studies}

Table \ref{tab:falcon3_audio_ablation} presents the results of our ablation studies, conducted on the Falcon3-Audio 7B model using the MMAU Benchmark test-mini Split. Full-test labels with 9000 samples for MMAU were not publicly available and required submissions to an external site with daily limits; therefore, we used the MMAU test-mini split with 1000 samples for all ablations. Moreover, a Spearman’s $\rho$ of 0.984 between test-mini and the test performance of different models in the current leaderboard confirms its suitability as a single, generalizable optimization metric without relying on paid LLM judges.

The following studies systematically analyzed the impact of various design choices.  We report the relative change in accuracy compared to the base Falcon3-Audio 7B model, which achieved a total accuracy of 65.0\%, with task-wise accuracies of 69.4\% (sound), 62.3\% (music), and 63.0\% (speech).

\textbf{Audio Encoder Variants:} We evaluated different sizes and pre-training data variations of the Whisper audio encoder.  Using the Whisper Medium (English) model provided the best performance. Larger (Large V3, Turbo) and multilingual (Medium - Multilingual) variants showed decreased performance, likely due to reduced specialization on English audio, which is dominant in the MMAU benchmark speech category.

\textbf{Audio Sequence Length Reduction:} We tested different stacking factors (2, 4, and 8) to reduce the audio sequence length.  Surprisingly, stacking either did not improve or slightly degraded performance, suggesting that aggressive sequence length reduction is not necessary for our architecture.

\textbf{Connector Architecture:} We investigated aggregating intermediate features from the Whisper encoder, both before and after the projection module. Neither approach yielded improvements, indicating that using only the final-layer features of Whisper is sufficient.

\textbf{Data Mixture:} We explored adding LibriSpeech (a 10K-hour subset) and the Auto-ACD dataset, individually and combined, to the training data. These additions did not improve performance, suggesting that the Open-ASQA and synthetic voice instruction dataset provide sufficient diversity and scale.

\textbf{Training Strategy:} We compared our single-stage training approach to a two-stage curriculum learning approach and found that the single-stage approach performed better.  We also experimented with freezing different modules during training and found that training all modules jointly (audio encoder, projector, and LLM) yielded the best results.  Freezing the LLM massively degraded performance.

\textbf{Hyperparameter Variants:} We evaluated the DoRA \cite{liu2024dora} variant of Low-Rank Adapatation and observed lower performance.

\section{Hyperparameters \& Prompt Format} 
\label{sec:hyperparameters}

\begin{figure}[H]
\centering
\includegraphics[width=0.7\columnwidth]{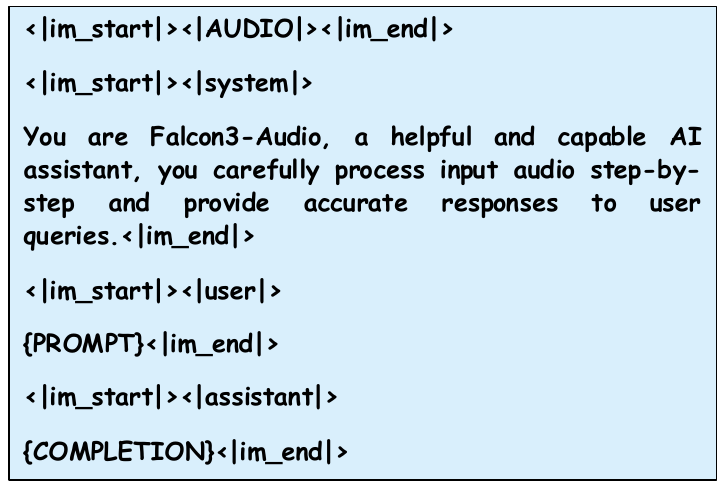}
\caption{Prompt template used for multi-modal instruction fine-tuning. The \texttt{<|AUDIO|>} token is replaced with audio features during fine-tuning.}
\label{fig:prompt_template}
\end{figure}

Table \ref{tab:hyperparameters} summarizes the key hyperparameters used for training the Falcon3-Audio models. Details on the training prompt template is shown in Figure~\ref{fig:prompt_template}.

\begin{table}[H]
\centering
\caption{Hyperparameters configuration for Falcon3-Audio.}
\begin{tabular}{ll}
\toprule
\textbf{Hyperparameter} & \textbf{Value}\\
\midrule
Epochs & 1 \\
Data Type & BF16 \\
Maximum Sequence Length & 4096 \\
Per Device Train Batch Size & 4 \\
Global Train Batch Size & 256 \\
Gradient Accumulation Steps & 1 \\
Maximum Learning Rate & 2e-4 \\
Learning Rate Scheduler & Cosine \\
Warmup Ratio & 0.01 \\
Weight Decay & 0.01 \\
Maximum Gradient Norm & 1.0 \\
LoRA Rank & 8 \\
LoRA Alpha & 16 \\
LoRA Dropout & 0.05 \\
Optimizer & AdamW \\
Audio Padding & Max Length \\
Global Padding Type & Longest \\
Global Padding Side & Right \\
\bottomrule
\end{tabular}
\label{tab:hyperparameters}
\end{table}

\section{Evaluation Benchmarks}
\label{sec:benchmarks}

\subsection{MMAU: Multi-Domain Audio Understanding}
The Multimodal Audio Understanding (MMAU) benchmark \cite{sakshi2024mmau} provides a rigorous evaluation of Audio Language Models (ALMs) across diverse audio domains, including speech, sound events, and music.  Comprising over 10,000 curated audio clips paired with multiple-choice questions, MMAU challenges models with 27 distinct skills requiring advanced auditory perception, contextual understanding, and reasoning.  The benchmark's test split includes a smaller ``test-mini'' set (1,000 samples) and a larger ``test'' set (9,000 samples).  Performance is measured using accuracy, providing a direct metric of a model's ability to correctly answer audio-based questions.  As the MMAU authors note, it is designed to be a particularly challenging benchmark, emphasizing multi-domain reasoning and comprehensive skill assessment for ALMs.

\subsection{AIR-Bench Foundational: Task-Specific Audio Understanding}
The AIR-Bench Foundational Benchmark evaluates ALMs through 19 distinct tasks encompassing around 24,683 single-choice questions. The benchmark spans diverse audio categories including speech, music, and environmental sounds, testing capabilities such as emotion recognition, intent classification, and question-answering for both sound and music domains. The evaluation framework uses a standardized single-choice Q\&A format, where the prompt is constructed by combining a question \emph{q} with candidate choices \emph{C}. While GPT-4 generates most questions, some QA tasks utilize questions directly from their source datasets. Similarly, candidate choices are either sourced from original datasets or generated via GPT-4, depending on task requirements. The evaluation process compares model-generated responses against reference answers, yielding binary scores that are averaged across all tasks to produce the final performance metric.



\subsection{AIR-Bench Chat: Open-Ended Audio Question Answering}
The AIR-Bench Chat Benchmark (Audio Instruction Benchmark) \cite{yang2024airbenchbenchmarkinglargeaudiolanguage} focuses on evaluating ALMs in open-ended, conversational scenarios.  It assesses performance across various audio types – speech, environmental sounds, music, and mixed audio – using around 2,200 question-answer pairs.  AIR-Bench uses GPT-4 Turbo (\emph{gpt-4-0125-preview} version) as a robust evaluator.  The evaluation process involves prompting GPT-4 Turbo to score model responses against GPT-4 generated reference answers, considering usefulness, relevance, accuracy, and comprehensiveness on a scale of 1 to 10. To mitigate potential position bias, the evaluator prompt is presented twice with swapped answer order, and the final score is the average of the two evaluations. This rigorous, GPT-4 based evaluation provides a nuanced assessment of ALM chat capabilities.

\section{Limitations of the Original AIR-Bench Foundational Evaluation}
\label{sec:airbench_limitations}

The original AIR-Bench evaluation code \cite{yang2024airbenchbenchmarkinglargeaudiolanguage} has limitations that can lead to inaccurate accuracy scores.  We illustrate this with an example:

\begin{quote}
\textbf{Question:} What instrument is most prominent in this piece? \\
A) drums \\
B) voice \\
C) electric guitar \\
D) bass \\
\textbf{Ground Truth:} C) electric guitar
\end{quote}

The original code relies on simply matching option IDs (e.g., ``A'', ``B'', ``C'', ``D'') at the beginning or end of the response, leading to three flaws:

\begin{enumerate}
    \item \textbf{Ignoring Correct Answers without IDs:}  A correct answer text (``electric guitar'') without the option identifier (``C)'') is incorrectly ignored.
    \item \textbf{Accepting Incorrect Answers with Correct IDs:} An incorrect answer with the correct option ID (e.g., ``C) drums") is incorrectly accepted.
    \item \textbf{Ignoring Non-Generations:} Empty responses (\emph{no generation}) are ignored, rather than treated as incorrect.
\end{enumerate}

Our modified evaluation script addresses these issues by checking response content, ensuring a more accurate and robust evaluation. For example, our script correctly identifies ``electric guitar'' as correct and ``C) drums'' as incorrect, regardless of ID presence.

\section{Qualitative Examples from AIR-Bench Chat and Potential Evaluator Bias}
\label{sec:airbench_chat_examples}

We observed that AIR-Bench Chat evaluations, using GPT-4 as a judge, can be sensitive to response style. Falcon3-Audio's factually correct but concise answers sometimes receive lower scores than the more verbose or stylistically similar responses of Qwen2-Audio Instruct. We present two illustrative examples below.


\begin{quote}
\textbf{Question:} What is the general mood of the piece? \\
\textbf{Ground Truth:} Positive and happy.

\textbf{Qwen2-Audio Instruct (Score: 10):} The piece has an uplifting and positive mood.

\textbf{Falcon3-Audio-7B (Score: 6):} The general mood of the piece is grand and majestic.
\end{quote}

\emph{Discussion:} Although ``grand and majestic'' could be consistent with ``positive and happy'' (depending on the audio), GPT-4 gives it a much lower score, suggesting a potential preference for responses closer in wording to the ground truth or Qwen2-Audio Instruct's style.








\begin{quote}
\textbf{Question:} What did the young male mention was originally between the platforms?

\textbf{Ground Truth:} An island platform.

\textbf{Qwen2-Audio Instruct (Score: 9):} An island platform was originally standing between them.

\textbf{Falcon3-Audio-7B (Score: 8):} The young male mentioned that an island platform stood in between them.
\end{quote}

\emph{Discussion:} Both responses convey the same correct information. The slightly lower score for Falcon3-Audio might be due to its more conversational phrasing.

\end{document}